# Recommendations for the Astronomy Graduate Admissions Process

*From the AAS Working Group on Graduate Admissions: Emily M. Levesque (chair), Courtney Dressing, Rachel Ivie, Grace Krahm, Meredith MacGregor, Daniel Piacitelli, Tom Rice*

Context: The AAS Board of Trustees thanks the AAS Working Group on Graduate Admissions (WGGA) for its work in developing this report. The Board has reviewed the report of the Graduate Admissions Task Force and has authorized the WGGA to share its recommendations, although these recommendations do not (as yet) represent an officially endorsed position of the AAS.

The AAS Working Group on Graduate Admissions (WGGA) would like to share the following recommendations for improving and standardizing key elements of graduate admissions in astronomy. Most astronomy graduate programs have large areas of overlap in their admissions processes; however, the existing small variations in requirements and mismatches in communication and transparency make admissions more challenging for students and programs alike. To improve this situation, and building on the work presented in the AAS Graduate Admissions Task Force (GATF) report [1] we recommend a few simple and straightforward changes for timelines, communication, and application content.

## Application Content

*We recommend that programs adopt an application format that consists of four key elements*:

1) Two 500-word[1] recommendation letters
2) One 1500-word application essay
3) Applicant CV
4) Unofficial transcripts

Below we provide some context for each recommendation.

**1) Two 500-word recommendation letters:** the GATF report found that recommendation letters in particular were mentioned by graduate programs as both an important and problematic part of the application review process. 97% of programs described recommendation letters as an "important" or "very important" part of the review process, and 82% required three letters (while the remaining 18% required two). However, programs also raised concerns about letter "inflation" (with recommenders writing increasingly-superlative and lengthy letters that make them difficult to interpret) and worries that the fame or familiarity of a letter writer might be unfairly weighted. There was also significant criticism of Likert-style ranking buttons (asking writers, for example, to rank individual traits or state whether an applicant is in the top 1%, 5%, 10%, etc. of students they have worked with); these are often required by letter submission interfaces despite overwhelmingly negative feedback on their usefulness. "Teaching letters" – where applicants request one of their three letters from professors who know them only through relatively impersonal coursework rather than through research or one-on-one interactions – were also cited as particularly ineffectual, both by programs and in an open letter by early career faculty [2]. Finally, programs noted that they would prefer shorter or more tightly-focused letters.

[1] Word counts rather than page counts are a much simpler and more straightforward way to establish and enforce length limits. This avoids any ambiguity introduced by font choice, font size, margin size, line spacing, images, etc. and discourages writers and readers alike from focusing on aesthetics rather than content.

We propose that programs require two 500-word recommendation letters for each applicant rather than three. This decreases both the number of letters (with the expectation that this will significantly decrease students' reliance on "teaching letters" in particular) and their length (500 words corresponds to approximately one full page of single-space 12-point text, or about a page and a half of text on university letterhead).

We also recommend that, whenever possible, recommendations consist solely of these letters with minimal reliance on online forms; these forms tend to include information that is either duplicated within the letter (for example, how long the writer has known the applicant) or largely seen as superfluous (for example, Likert-style rankings) and are an extremely poor use of letter writers' time. The WGGA recognizes that following this advice in particular may be out of programs' hands given that most letters are submitted through college or university interfaces that include form answers and rankings as required content. In such cases, we recommend that programs include clear language for letter writers (and admissions committee members) detailing how such content will be used in the evaluation process.

**2) One 1500-word application essay:** the GATF report found that 84% of graduate programs describe applicant essays as "important" or "very important". However, there is currently significant variation in how programs describe their essay requirements: these tend to be one or two statements, with a broad variety of page-based lengths and hard-to-interpret descriptive titles (e.g., "applicant statement", "statement of purpose", "personal statement", "academic statement", "research statement"). Applicants also cited application essays and their corresponding workload – particularly the task of writing a single essay and then tailoring it repeatedly to match a broad variety of formatting requirements – as one of the most challenging parts of the application process.

A standard number (1) and length (1500 words, or approximately 3 pages) for essays will substantially reduce the burden on applicants to write and rewrite very similar content in a variety of different formations for each program. Even a minimal template for this essay would also help to standardize and clarify the key elements most programs require: for example, an essay broken down into 1200 words encompassing the content currently contained in research and personal statements, followed by 300 words in which applicants clearly detail their interest in a specific program (put plainly, this serves as a formalization of the "tailoring paragraph" format that is already often adopted by applicants). A single widely-adopted format would make it easy for applicants to write a main application essay and then modify it for different programs, and would also make it easy for admissions committees to equitably evaluate this written component of their applications by clarifying where applicants will address their preparedness, their personal background, and their interest in a program. This template could be as simple as the word-count breakdown described above; it could also take the form of a simple LaTeX/document template suggested by the AAS.

**3) Unofficial transcripts:** at the initial application review process unofficial transcripts should be requested if at all possible. Official transcripts are logistically and financially burdensome for applicants to request and, if need be, can be requested from shortlisted or admitted students at a later point in the process.

**4) CV:** conventionally this format is largely left up to students, but it is helpful to make clear that these are typically no more than 1-2 pages. A template or example could also be provided to offer broad content and format guidance.

We strongly discourage “optional” content. Application materials explicitly described as “optional” are prone to misinterpretation by both applicants and admissions committees, often implying that not including such materials might be interpreted as a red flag or de facto penalty. At the same time, explicitly banning applicants from submitting specific information (such as, for example, GRE scores or research materials) is also not recommended; any strong policy along these lines must come with the ability and willingness to enforce it, which takes administrative effort and can be inequitably applied. Rather than dealing with optional or forbidden content, programs should simply make clear, to both applicants and admissions committees, what the application requirements are and how any unrequested content will be used.

We also recommend that, where possible, programs either eliminate application fees or – as the administration of these fees is often beyond the control of individual departments – make information on any fee waivers that might be available as easy to find as possible (including waiver deadlines, criteria, and information on how to request them). Applying to graduate school is extremely expensive for most students: the GATF report found that prospective students applied to 11-12 programs on average, and that the median application fee is $75. A typical applicant therefore spends ~$900 applying to graduate school. Along with the volume of the workload and a lack of process transparency, the financial burden of admissions was also listed by applicants as one of the most challenging parts of the admissions process.

**<u>Timeline and Communication</u>**

**1) April 1st “down-select” date:** *we strongly encourage applicants with early offers from multiple PhD programs to narrow their choice down to their top two programs by April 1st.* Declining offers in a timely fashion (rather than in the final few days of the admissions process) allows programs to extend offers to new prospective students, gives those students more time to make their decisions, and supports better management of program yield and planned class size. The GATF report found that between 2018-2023, 45% of applicants who received offers received them from 3 or more schools; this represents a significant number of offers that will eventually be declined. When possible, doing so by April 1st increases the likelihood of these offers being successfully extended to other waiting applicants.

We emphasize that this is a recommendation and *not* an official or enforceable deadline. Per the “April 15th Resolution” agreed to by the Council of Graduate Schools, “[s]tudents are under no obligation to respond to offers of financial support prior to April 15; earlier deadlines for acceptance of such offers violate the intent of this Resolution” [3]. Applicants should take all the time they need and make full use of the information offered to them by prospective programs, advisers, peers, and other public resources (for example, the Astrobites guide to choosing a graduate school [4]) when making this important choice. Our intent is for the April 1st down-select date to serve as a guide for the decision-making process and a target date for applicants who are considering multiple offers.

**2) Program transparency and communication:** *we strongly encourage graduate programs to clearly communicate dates, decisions, and updates on their admissions process.* In particular, whenever possible, programs should commit to and share key dates (via website updates, emails, or similar) when they will share news with applicants.

The GATF report highlighted poor communication and opaque timeline information as a significant source of stress for applicants, who noted that ambiguity on when they would receive application decisions or updates was especially difficult. This uncertainty can place enormous pressure on programs and applicants alike to make critical decisions about offers and acceptances with incomplete information or on extremely short timescales.

Clearly-stated “notification dates” from programs, for example, would inform applicants that they can expect concrete news on their application — admission, rejection, or placement on a waitlist — by a specific date (it is, of course, crucial that programs are able to then follow through on this promise). This gives applicants a clearer and less stressful timeline for when they can expect news, and when they will have a complete set of information regarding their grad school choices and can proceed to the decision-making stage. The WGGA also encourages programs to send rejections as soon as candidates are removed from future selection rather than waiting to send acceptance and rejection letters simultaneously.

The WGGA recognizes that not all decision timelines can be easily predicted, particularly at later stages in the application process (for example, offers from waitlists). However, even committing to regular check-in dates where applicants’ status is confirmed or updated can keep lines of communication open and increase transparency for students.

Finally, it is worth noting that one benefit of the recommendations we describe above – making the process more streamlined and affordable for applicants – carries with it a significant potential challenge: this could lead to an increase in the number of applications that each student submits and the size of programs’ applicant pools. However, it is important to note that limiting the number of submitted graduate school applications based on students’ time, money, and resources is not an effective or equitable way to manage applicant pool sizes since it risks discouraging promising students that cannot currently meet the demands of a needlessly inefficient process.

Our hope is that some of these effects can be mitigated by the standardization process itself: decreasing the number and length of letters, limited applications to a single 3-page essay, and following even a minimal template for these essays should make the reading and reviewing process easier for committees. Timeline recommendations can also help keep workloads manageable, particularly on the decision end of the process. However, the AAS WGGA recognizes that the growth of applicant pools is an ongoing challenge in the field and will require further attention.